\documentclass[useAMS,usenatbib,referee]{mn2e}

\usepackage{epsfig,amsmath,amssymb,amsfonts,mathrsfs,latexsym,graphicx,lscape}
\usepackage{times}
\usepackage{longtable}
\usepackage{xcolor}

\begin{document}

\LARGE \normalsize \title[LFQPOs in GRS 1915+105]{A Statistical Study on the Low-Frequency QPO Amplitude Spectrum and Amplitude in GRS~1915+105}

\author[Yan et al.]
       {Shu-Ping Yan$^{1,2}$\thanks{E-mail: yanshup@xao.ac.cn}, Guo-Qiang Ding$^{1}$, Na Wang$^{1}$\thanks{E-mail: na.wang@xao.ac.cn}, Jin-Lu Qu$^{3}$, Li-Ming Song$^{3}$\\
       \newauthor  \\ 
      $^{1}$Xinjiang Astronomical Observatory, Chinese Academy of Sciences, 150, Science 1-Street, Urumqi, Xinjiang 830011, China\\
      $^{2}$University of Chinese Academy of Sciences, 19A Yuquan road, Beijing 100049, China\\
      $^{3}$Key Laboratory for Particle Astrophysics, Institute of High Energy Physics, Chinese Academy of Sciences, 19B Yuquan Road, Beijing 100049, China}

\date{Accepted 2013 June 3.  Received 2013 June 3; in original form 2013 January 29}
\pubyear{2013}
\pagerange{\pageref{firstpage}--\pageref{lastpage}}

\maketitle
\label{firstpage}

\begin{abstract}

A statistical study was made on both the energy dependence of the low-frequency quasi-periodic oscillation (QPO) amplitude (LFQPO amplitude spectrum) and the LFQPO amplitude from all the {\it RXTE} observations of GRS~1915+105. Based on the two-branch correlation of the LFQPO frequency and the hardness ratio, the observations that were suitable for evaluating the LFQPO amplitude spectrum were divided into two groups. According to a comparison between the radio and X-ray emissions, we deduced that the jets during the two groups of observations are very different. A negative correlation between the LFQPO frequency and the radio flux was found for one group. The LFQPO amplitude spectrum was fitted by a power-law with an exponential cutoff in order to describe it quantitatively. It reveals that as the LFQPO frequency increases, the power-law hardens. And the cutoff energy firstly decreases, and then smoothly levels off. The fit also shows that the LFQPO amplitude spectra of the two groups are essentially the same, suggesting that the LFQPO seems not to originate from the jet. The LFQPO amplitude spectra are hard, indicating a possible origin of the LFQPO in the corona. As the LFQPO frequency increases, the LFQPO amplitude firstly increases and then decreases. The effects of the low pass filter and the jet on the LFQPO amplitude are discussed.

\end{abstract}

\begin{keywords}
accretion, accretion disks -- black hole physics -- X-rays: individuals (GRS~1915+105) -- X-rays: binaries
\end{keywords}

\section{Introduction}

The black hole binary system (BHB) GRS~1915+105 was discovered in 1992 with WATCH on board {\emph{GRANAT}} \citep{Castro92}. It is at a distance of $\sim 11$ kpc away \citep[e.g.][]{Fender99, Zdziarski05}, and comprises a spinning black hole \citep {Zhang97, McClintock06} with mass $14\pm4$ $M_{\odot}$, and a K-M III giant star with mass $0.8\pm0.5$ $M_{\odot}$ as the donor \citep{Harlaftis04, Greiner01b}. The orbital separation of the binary components is about $108\pm4$ $R_{\odot}$, and the orbital period is $33.5\pm1.5$ d \citep{Greiner01a}. GRS 1915+105 was the first microquasar to be found and produces superluminal radio jets \citep{Mirabel94, Fender99}. 

It shows various X-ray light curves and complex timing phenomena. Based on the appearances of light curves and color-color diagrams, the behaviours of GRS~1915+105 are classified into more than ten categories \citep{Belloni00, Klein02, Hannikainen05}. These categories can be reduced to transitions among three basic states, namely states A, B, and C. Three types of quasi-periodic oscillations (QPOs) with different QPO frequency bands have been observed in GRS~1915+105 \citep[e.g.][]{Morgan97, Chen97, Strohmayer01, Belloni01, Belloni06}. The low-frequency ($\sim$ 0.5--10 Hz) QPO (LFQPO) is the type most commonly observed. Considerable effort has been put into exploring the origin of the LFQPO of GRS~1915+105. It has been shown that the LFQPO frequency is positively correlated with the fluxes of the thermal and power-law components as well as with the total flux \citep[e.g.][]{Chen97, Markwardt99, Muno99, Trudolyubov99, Reig00, Tomsick01, Muno01}. \citet{Muno99} and \citet{Rodriguez02b} reported that as the LFQPO frequency increases, the temperature of the inner accretion disk increases, and the disk radius decreases. These results indicate that the LFQPO is related to both the accretion disk and the region where the hard component is produced, e.g. corona. It should be noted, however, that most of these results are dependent on spectral models, but the origin of the hard spectral component is still a matter of debate \citep[e.g.][]{Muno99, Rau03, Vadawale03, Zdziarski05, Titarchuk09, Oers10, Neilsen11}. 

As model-independent approaches, it is useful to study the LFQPO frequency-LFQPO amplitude relationship, the energy dependence of the low-frequency QPO frequency (LFQPO frequency spectrum), and the energy dependence of the low-frequency QPO amplitude (LFQPO amplitude  spectrum) for GRS~1915+105. The LFQPO amplitude refers to the LFQPO fractional rms amplitude which is measured by using a Lorentzian fit to the power spectrum (see Section 2 for details). It was found that the LFQPO amplitude is inversely correlated with the LFQPO frequency \citep[e.g.][]{Muno99, Trudolyubov99, Reig00}. \citet{Qu10} studied the LFQPO frequency spectrum of GRS~1915+105 and found that as the centroid frequency of the LFQPO increases, the relationship between LFQPO frequency and photon energy evolves from a negative correlation to a positive one. Three additional combined patterns of the negative correlation and the positive one were discovered \citep{Yan12}. Besides, as photon energy increases, the LFQPO amplitude increases and then flattens in some cases \citep[e.g.][]{Tomsick01, Rodriguez02a, Rodriguez04, Zdziarski05, Sobolewska06}, indicating a possible association between the LFQPO and the corona \citep[e.g.][]{Morgan97, Ingram12}.

Nevertheless, there is no statistical study in which all the {\it RXTE} observations of GRS~1915+105 have been utilized on both the LFQPO amplitude spectrum and the LFQPO amplitude. In order to reveal more details of the LFQPO phenomenology and to investigate the origin of the LFQPO, in this study we analysed all the {\it RXTE}/PCA data of GRS~1915+105 and found that as the LFQPO frequency increases, the LFQPO amplitude spectrum becomes harder, and the LFQPO frequency-amplitude relationship evolves smoothly from the positive correlation to the negative one. A negative correlation between the LFQPO frequency and the radio flux is also found. The observations and data reduction methods are described in Section 2, the results are presented in Section 3, and a discussion and the conclusions are given in Section 4.

\section{Observations and Data Reduction}

We searched the LFQPO from all the {\it RXTE}/PCA observations of GRS~1915+105. Only some observations are suitable for evaluating the LFQPO amplitude spectrum. The LFQPO frequency sometimes varies obviously during an observation. For obtaining credible result, this kind of observation is split into several time intervals when the LFQPO frequency is relatively stable. A total of 168 observation intervals when the X-ray emission is relatively hard and steady are obtained (Table \ref{table1}).

It is a common technique to combine the timing analysis with the spectral analysis. In view of the debate on spectral model, we investigate the relationship between the hardness ratio (HR) and the LFQPO frequency as an approximate spectral analysis. The HR is defined as the ratio of the count rate in 7--60 keV to that in 2--7 keV. The corresponding PCA absolute channel intervals of the two energy bands in PCA gain epochs 3, 4, and 5 are 19--255 and 0--18,  16--255 and 0--15, as well as 17--255 and 0--16, respectively. The count rate is obtained by extracting the background-subtracted PCA Standard-2 light curve using the HEASOFT version 6.7 package.

For investigating the LFQPO amplitude spectrum, the light curves are extracted from the binned and event mode data. Good time intervals are defined as follows: a satellite elevation over the Earth limb $>10^{\circ}$ and an offset pointing $<0.02^{\circ}$. In order to acquire the details of the LFQPO amplitude spectrum with high enough confidence, only the binned mode data with energy channel number $\geq 4$ and time resolution $\leq 8$ ms are selected. The light curves are extracted with a time resolution of 8 ms in PCA energy bands defined in Table S1 in the Supporting Information. The power density spectra (PDSs) are computed on 64-s sampling duration, with the normalization of \citet{Miyamoto92}, which gives the periodogram in units of (rms/mean)$^2$/Hz. The Poisson noise is also corrected \citep[e.g.][]{Klis89, Vaughan03}. Following \citet{Belloni02}, we fit the PDS with a model that includes several Lorentzians to represent the LFQPOs, the continuum, and other broad features, respectively. The uncorrected LFQPO amplitude is defined as $A_{\rm raw} (\%rms) = 100 \times \sqrt{WN\pi/2}$, where $W$ is the full width at half maximum (FWHM) of the Lorentzian that represents the LFQPO and $N$ is the Miyamoto normalization of the Lorentzian. The LFQPO amplitude is further corrected for background \citep{Berger94, Rodriguez11}. The errors are derived by varying the parameters until $\Delta\chi^2=1$, at $1\sigma$ level.

For studying the LFQPO frequency-amplitude relationship, the light curves are extracted from the binned mode data in PCA absolute channel 0--35 ($\sim$ 2--13 keV) and the event mode data in channel 36--255 ($\sim$ 13--60 keV) with a time resolution of 8 ms. With the asynchronous rows being deleted from the FITS files, the binned and event mode light curves in the same observation interval are added together to obtain a light curve that will be used to measure the LFQPO frequency and amplitude.

\section{Results}

\subsection{LFQPO Frequency-Hardness Ratio Relationship}

\begin{figure*}
\centerline{\includegraphics[height=9cm,angle=-90]{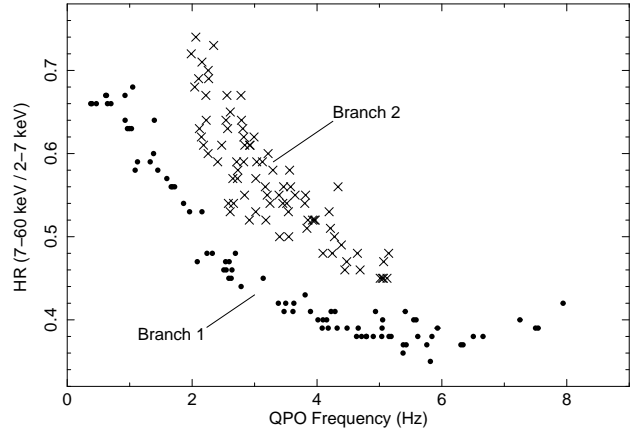}}
\caption{The hardness (HR) as a function of the LFQPO frequency. The points form two obviously separated branches termed Branch 1 and Branch 2, respectively. The Branch 1 observations are marked with filled circles and the Branch 2 observations are marked with crosses, and similarly in subsequent figures.}
\label{fig:fre2rate2HR} 
\end{figure*}

Fig. \ref{fig:fre2rate2HR} shows the relationship between the LFQPO frequency and the HR. One can see that the points in this figure form two obviously separated branches. In order to clearly describe and analyse the results, we refer to the lower branch as ``Branch 1'' (filled circles) and the upper branch as ``Branch 2'' (crosses), and divide these observation intervals into two groups corresponding to the two branches, respectively. Branch 1 is in the range $\sim$ 0.4--8 Hz and Branch 2 is in the range $\sim$ 2--5.5 Hz. For Branch 1, as the LFQPO frequency increases, the HR firstly decreases, then smoothly levels off, and then increases slightly. For Branch 2, the HR decreases over the entire range.

\subsection{Spectral States of the Branch 1 and Branch 2 Observations}

\begin{figure*}
\centering
\centerline{\includegraphics[height=16.8cm,angle=-90]{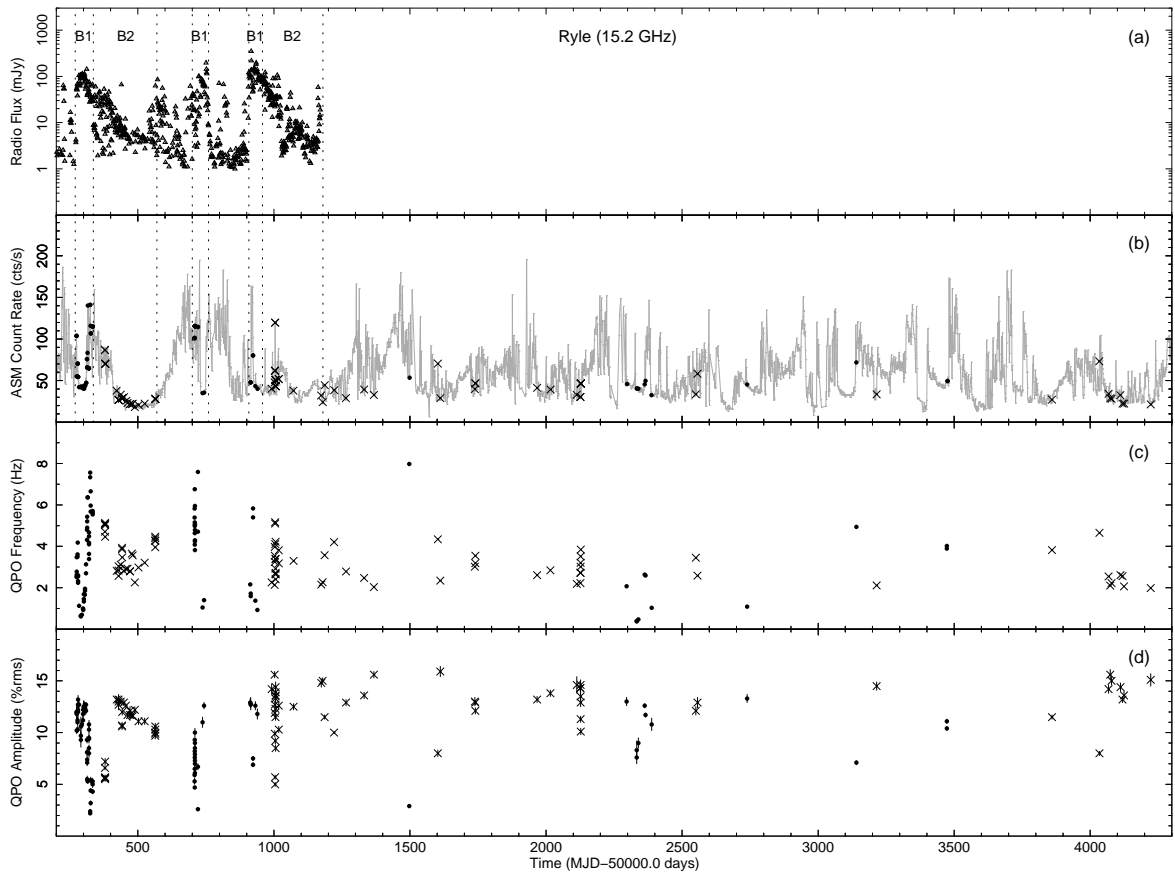}}
\caption{(a) Flux as a function of time from the Ryle Telescope at 15.2 GHz. The radio fluxes are obtained from Fig. 1 in \citet{Muno01}. The vertical dotted lines are plotted to show clearly the radio conditions of the Branch 1 and Branch 2 observations. (b) The {\it RXTE}/ASM light curve (gray curve) and the observation times of the two groups of observations. The bin size is 1 d. The data are provided by the ASM/{\it RXTE} teams at MIT and at the {\it RXTE} Science Operations Facility and Guest Observer Facility at NASA's Goddard Space Flight Center. (c) The LFQPO frequency and (d) amplitude as functions of the time.} 
\label{fig:ASMbranch}
\end{figure*}

\citet{Muno01} investigated the radio and X-ray properties of GRS~1915+105 when its X-ray emission is hard and steady, and defined three spectral states/conditions. The energy spectra of the Branch 1 and Branch 2 observations are different based on the two-branch correlation of the LFQPO frequency and the HR. Then, it is useful to identify the spectral states of the Branch 1 and Branch 2 observations. In order to make clear the states of the two groups of observations, we plot the {\it RXTE}/ASM count rate and the radio flux from the Ryle Telescope at 15.2 GHz as functions of time, and show the times of the observations analyzed in this work (Fig. \ref{fig:ASMbranch}a,b). The values of the radio flux were obtained from \citet{Muno01}. At first glance, the Branch 1 observations are in the time intervals (B1s in Fig. \ref{fig:ASMbranch}a) when GRS~1915+105 produces the brightest radio emissions, and the Branch 2 observations are in the time intervals (B2s in Fig. \ref{fig:ASMbranch}a) when GRS~1915+105 produces fainter radio emissions.

The LFQPO frequency and amplitude as functions of time are also presented. The behaviour of the LFQPO frequency is somewhat similar to the behaviour of the count rate (Fig. \ref{fig:ASMbranch}c). The LFQPO amplitude is, however, a non-monotonic function of the count rate (Fig. \ref{fig:ASMbranch}d). 

\begin{figure*}
\centering
\centerline{\includegraphics[height=8cm,angle=-90]{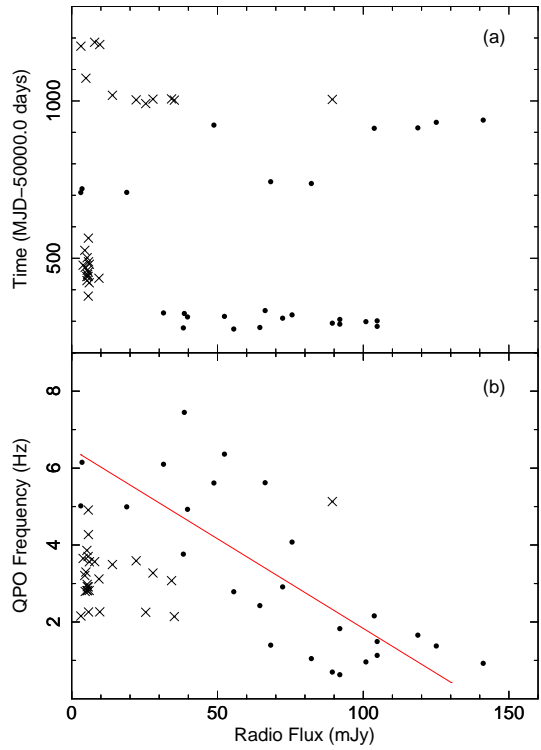}}
\caption{The observations shown in Fig. \ref{fig:ASMbranch} are binned into time intervals. The bin size is 1 day. (a) The radio flux as a function of time. (b) The relationship between the radio flux and the LFQPO frequency.} 
\label{fig:rad2tim2fre}
\end{figure*}

In order to show the relationship between the radio emission and the LFQPO more clearly, we need to bin the observations into time intervals. The time intervals of the radio fluxes presented in Fig. \ref{fig:ASMbranch} are about 1 d. We therefore select 1 d as the bin size. Fig. \ref{fig:rad2tim2fre} shows the radio flux as a function of time, and the relationship between the radio flux and the LFQPO frequency. Obviously, most of the radio fluxes corresponding to Branch 1 observations are larger than 30 mJy, and all except one fluxes corresponding to Branch 2 observations are lower than 40 mJy. As shown in Fig. \ref{fig:rad2tim2fre}(b), for Branch 1, the LFQPO frequency is negatively correlated with the radio flux. For Branch 2, it has no obvious correlation with the radio flux. The points of Branch 1 in Fig. \ref{fig:rad2tim2fre}(b) are fitted using least squares. The slope of the best-fitting line is $-0.047\pm0.015$ Hz mJy$^{-1}$, and the adjusted R$^2$ is 0.62.

The Branch 2 point whose radio flux is about 90 mJy is located at some distance from the main Branch 2 group. We therefore have checked it and found that its {\it RXTE} observation time was a bit earlier than its Ryle observation time, and the radio fluxes on the days around this observation time was $\sim$ 20--30 mJy. It is thus possible that the radio flux corresponding this {\it RXTE} observation is not actually so high, and the outlying Branch 2 point might actually be located within the main group. This is just a speculation, however, owing to the lack of data. We also checked all the other radio fluxes and found that they are close to those on either side of them. These fluxes thus seem to be more reliable, although some uncertainty still exists. The states of the two groups of observations will be discussed in Section 4.

\subsection{LFQPO Amplitude Spectrum}

\begin{figure*}
\centerline{\includegraphics[height=17.8cm,angle=-90]{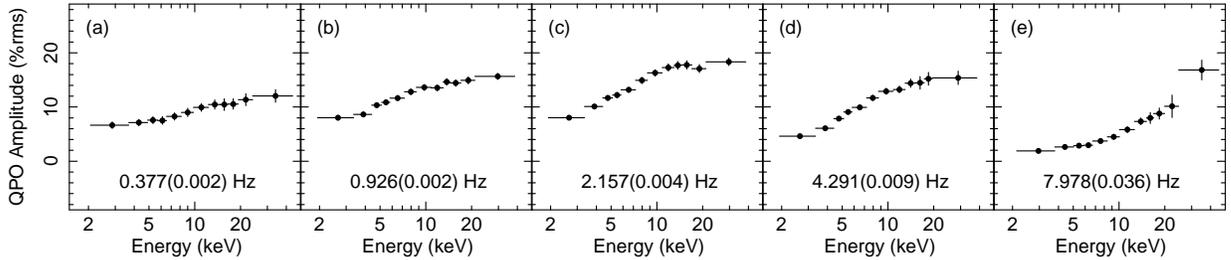}}
\caption{Representative LFQPO amplitude spectra of the Branch 1 observations. Their observation IDs are (a) 60701-01-16-00, (b) 30703-01-17-00, (c) 30402-01-09-01, (d) 20186-03-02-06d, and (e) 40703-01-38-02, respectively. The frequencies shown are the LFQPO centroid frequencies. The horizontal bars denotes the width of the energy band. The vertical bars are error bars.} 
\label{fig:B1}
\end{figure*}

\begin{figure*}
\centerline{\includegraphics[height=14.4cm,angle=-90]{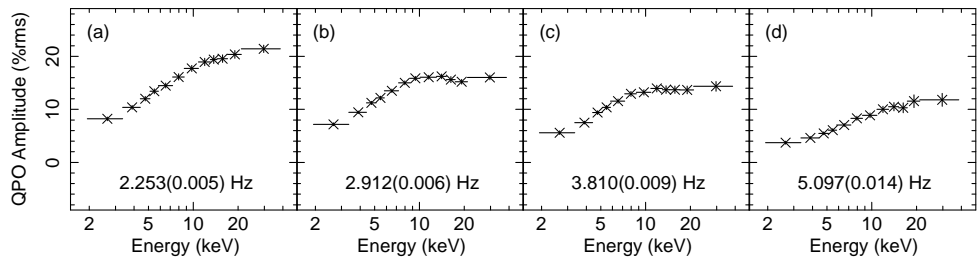}}
\caption{Representative LFQPO amplitude spectra of the Branch 2 observations. Their observation IDs are (a) 30703-01-22-00, (b) 20402-01-10-00, (c) 30703-01-25-00b, and (d) 30182-01-03-00a, respectively. The frequencies shown are the LFQPO centroid frequencies. The horizontal bars denotes the width of the energy band. The vertical bars are error bars.}
\label{fig:B2}
\end{figure*}

For each interval listed in Table \ref{table1}, we have drawn a diagram to show the LFQPO amplitude spectrum. Although these spectra have various shapes, they evolve with the LFQPO frequency. Fig. \ref{fig:B1} shows several representative spectra of the Branch 1 observations. When the LFQPO frequency is very low, the amplitude increases slightly with energy (Fig. \ref{fig:B1}a). As the LFQPO frequency increases, the amplitude in the higher energy band gradually increases (Fig. \ref{fig:B1}b and c), and then the amplitudes in both the higher and lower energy bands gradually decrease (Fig. \ref{fig:B1}d,e). For the LFQPOs with higher frequency, however, the amplitude in higher energy band is relatively high (Fig. \ref{fig:B1}e). Fig. \ref{fig:B2} shows representative spectra of the Branch 2 observations. When the LFQPO frequency is low, the amplitude spectrum is steep (Fig. \ref{fig:B2}a). As the LFQPO frequency increases, the amplitude in higher energy band decreases (Fig. \ref{fig:B2}b), and then the amplitudes in both the higher and lower energy bands decrease (Fig. \ref{fig:B2}c,d).

\begin{figure*}
\centerline{\includegraphics[height=8cm,angle=-90]{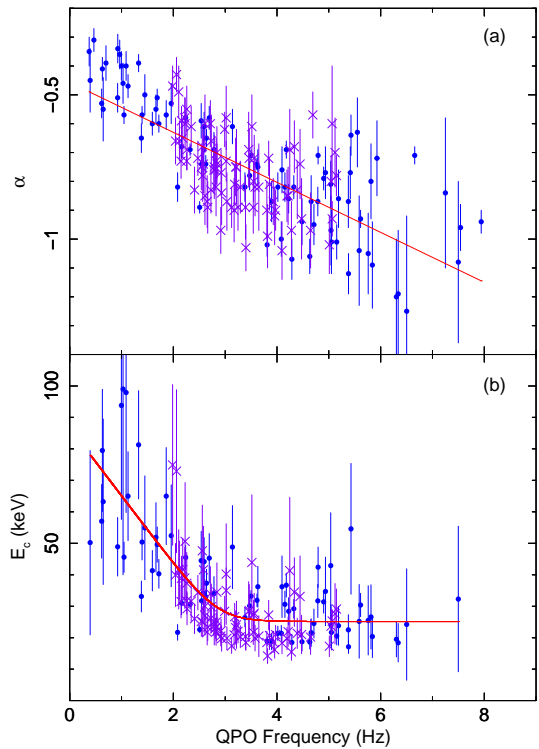}}
\caption{The LFQPO amplitude spectrum is fitted by a power-law with an exponential cutoff. (a) The power-law index as a function of the LFQPO frequency. The points are fitted with the least-squares. (b) The cutoff energy as a function of the LFQPO frequency. The points are fitted with function (1) in \citet{Shaposhnikov07}.}
\label{fig:fre2ga2cut}
\end{figure*}

In order to quantitatively describe the LFQPO amplitude spectrum, we fit the spectrum by a power-law with an exponential cutoff, $A(E)=KE^{-\alpha}\exp(-E/E_{\rm c})$, where $\alpha$ is the power-law index and $E_{\rm c}$ is the e-folding energy of exponential roll-off. Fig. \ref{fig:fre2ga2cut} shows $\alpha$ and $E_{\rm c}$ as functions of the LFQPO frequency. Clearly, the two group points are essentially identical, indicating that there is not much difference between the two groups of observations in the LFQPO amplitude spectrum. Thus, the two group points will later be fitted as a whole. As the LFQPO frequency increases from $\sim 0.4$ Hz to $\sim 8$ Hz, $\alpha$ decreases from $\sim -0.4$ to $\sim -1.1$. (Fig. \ref{fig:fre2ga2cut}a). The points are fitted with least squares. The slope of the best-fitting line is $-0.087\pm0.012$ Hz$^{-1}$, and the adjusted R$^2$ is 0.53. Fig. \ref{fig:fre2ga2cut}b presents the LFQPO frequency dependence of $E_{\rm c}$. At lower LFQPO frequencies the dependence starts at $E_{\rm c}$ of $\sim 80$ keV and follows a negative correlation untill a certain LFQPO frequency, where it levels off. The points are fitted with the function $E(f)=A-D\,B\ln\{ \exp [(f_{\rm tr}-f)/D]+1\}$ \citep[function (1) in][]{Shaposhnikov07}. The best-fitting values are $A=25.1\pm2.8$ keV, $B=-21.6\pm8.0$ Hz$^{-1}$, $D=0.3\pm0.5$ Hz and $f_{\rm tr}=2.84\pm0.55$ Hz. The errors for the best-fitting parameters are standard deviations.

\subsection{LFQPO Frequency-Amplitude Relationship}

\begin{figure*}
\centerline{\includegraphics[height=9cm,angle=-90]{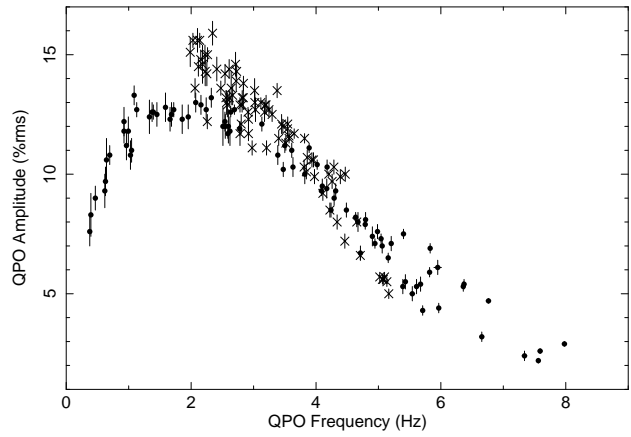}}
\caption{The relationship between the LFQPO frequency and the LFQPO fractional amplitude.}
\label{fig:fre2rms} 
\end{figure*}

The relationship between the LFQPO frequency and amplitude is shown in Fig. \ref{fig:fre2rms}. For Branch 1, as the LFQPO frequency increases, the LFQPO amplitude increases from $\sim 7\%$ to $\sim 13\%$ at $f < 2$ Hz, and then decreases from $\sim 13\%$ to $\sim 2\%$ at $f > 2$ Hz. For Branch 2, as the LFQPO frequency increases, the LFQPO amplitude decreases monotonically from $\sim 16\%$ to $\sim 5\%$. 

\begin{figure*}
\centerline{\includegraphics[height=9cm,angle=-90]{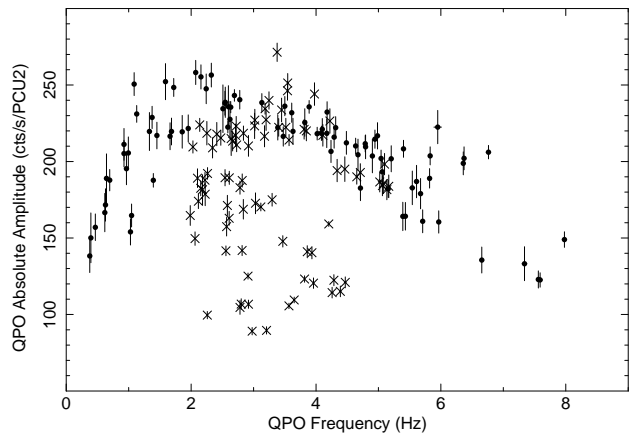}}
\caption{The relationship between the LFQPO frequency and the LFQPO absolute amplitude.}
\label{fig:rmsxrate} 
\end{figure*}

The LFQPO absolute amplitude is estimated by multiplying the LFQPO amplitude with the corresponding count rate \citep[see, e.g.][]{Mendez97, Gilfanov03, Zdziarski05}. For Branch 1, as the LFQPO frequency increases, the LFQPO amplitude increases at $f < 2$ Hz and then decreases, similar to the behaviour of the LFQPO amplitude. For Branch 2, the points are widely scattered (Fig. \ref{fig:rmsxrate}).

\section{Discussion and Conclusions}

LFQPOs have been detected in many BHBs \citep[see e.g.][]{Klis04, McClintock06, Remillard06}. 
Their frequencies and amplitudes are usually correlated with spectral parameters of both the thermal and power-law components \citep[e.g.][]{Chen97, Markwardt99, Muno99, Trudolyubov99, Reig00, Sobczak00, Revnivtsev00, Tomsick01, Muno01, Vignarca03}. However, neither the QPO mechanism \citep[e.g.][]{Stella98, Stella99, Wagoner99, Tagger99, Chakrabarti00, Nobili00, Titarchuk00, Psaltis00, Ingram09} nor the origin of the power-law component \citep[see, e.g.][]{Done07} is very clear. Thus, we take a model-independent strategy to study the phenomenon of the LFQPO. 

Based on the statistical study of both the LFQPO amplitude spectrum and the LFQPO amplitude of GRS~1915+105, we find that in the LFQPO frequency-HR diagram the points form two branches which are designated as Branch 1 and Branch 2 (Fig. \ref{fig:fre2rate2HR}). This indicates that the energy spectra of the observations corresponding to the two branches are very different. Similar phenomenon has also been found by other authors. For instance, \citet{Belloni00} showed that the $\chi$ state points follow two branches in the color-color diagram and used different spectral models for the observations located on the two branches, respectively. \citet{Rau03} studied four years of {\it RXTE} observations of GRS~1915+105 during the $\chi$ state and revealed a two-branch correlation of the power-law slope and the power-law normalization. Their two branches correspond to our Branch 1 and Branch 2. \citet{Oers10} analyzed two observations which belong to Branch 1 and Branch 2, respectively, and found that their best-fitting model parameters are significantly different.

Identifying the states of the Branch 1 and Branch 2 observations is helpful to analyze the properties and origin of the LFQPO. \citet{Muno01} investigated the radio and X-ray properties of GRS~1915+105 when the X-ray emission is hard and steady, and established that radio emission always accompanies the hard state of GRS~1915+105, but that the radio flux and the X-ray flux are not correlated. They defined ``radio plateau conditions" (the radio flux at 15.2 GHz $> 20$ mJy and the radio spectrum is optically thick with power law $E^{-\alpha_{\rm r}}$ where $\alpha_{\rm r} < 0.2$), ``radio steep conditions" (the radio flux at 15.2 GHz $> 20$ mJy and the radio spectrum is optically thin with $\alpha_{\rm r} > 0.2$), and ``radio faint conditions" (the radio flux at 15.2 GHz $< 20$ mJy) for the hard-steady X-ray observations. The radio emission is generally believed to be synchrotron emission from ejected plasma in sporadic or continuous jets \citep[e.g.][]{Fender95}. For GRS~1915+105, the optically thick radio emission during plateau conditions has been resolved as a compact jet of relativistic electrons \citep{Dhawan00}. The optically thin radio emission during steep conditions is originated from material ejected from the central source \citep{Mirabel94, Fender99, Dhawan00}. The radio faint observations show some properties similar to a weak radio plateau state. The radio steep observations represent the transition into and out of radio plateau conditions. As the LFQPO frequency decreases, the radio emission becomes brighter and optically thick. The source is in plateau conditions when the LFQPO frequency is less than 2 Hz. By combining our results showed in Section 3.2 with the definitions and conclusions in literature presented here, we deduce that for Branch 1, as the LFQPO frequency increases, the source evolves from the radio plateau conditions to the radio faint conditions via the radio steep conditions, and for Branch 2, the source are mainly in the radio faint conditions (Figs \ref{fig:ASMbranch} and \ref{fig:rad2tim2fre}). It is necessary to point out that the radio conditions of some observations cannot be identified due to the lack of radio data. Nevertheless, the two branches are clearly separated and smoothly evolve in the LFQPO frequency-HR diagram, which reflects the smooth evolution of spectral state. Therefore, the lack of radio data has no effect on our identification of state.

Despite the significant difference between the spectral states of the Branch 1 and Branch 2 observations, and the fact that the radio emissions at the same LFQPO frequency are always stronger during the Branch 1 observations than during the Branch 2 observations (Fig. \ref{fig:rad2tim2fre}b), there is no essential difference between the LFQPO amplitude spectra of the two branches (Figs \ref{fig:B1}, \ref{fig:B2} and \ref{fig:fre2ga2cut}). Thus, the LFQPO seems not to originate from the jet, as the jets of GRS~1915+105 during the Branch 1 and Branch 2 observations are very different. The spectrum of the LFQPO amplitude is hard (Figs \ref{fig:B1} and \ref{fig:B2}), suggesting that the LFQPO is related to the corona \citep[e.g.][]{Morgan97, Ingram12}. \citet{Shaposhnikov07} showed that as the LFQPO frequency increases, the spectral index of the X-ray spectrum, $\alpha_{\rm x}$, increases linearly and then smoothly levels off to become a constant. The $\alpha_{\rm x}$-LFQPO frequency relationship is fitted with function (1) in \citet{Shaposhnikov07}, and the obtained transition frequency is $2.23\pm0.07$ Hz. Coincidentally, we find that as the LFQPO frequency increases, $E_{\rm c}$ of the LFQPO amplitude spectrum also decreases and then smoothly levels off (Fig. \ref{fig:fre2ga2cut}b). The $E_{\rm c}$-LFQPO frequency relationship is fitted with the same function, and the transition frequency is $2.84\pm0.55$ Hz. The similarity of the behaviours of the two correlations is another indicator of the link between the LFQPO and the corona, although the details are not yet very clear.

For Branch 1, the LFQPO amplitude increases with frequency until $\sim 2$ Hz; above this, it decreases. Negative correlations between LFQPO amplitude and frequency have been observed in GRS~1915+105 and other BHBs \citep[e.g.][]{Muno99, Sobczak00, McClintock09, Heil11}. The aperiodic variability also shows a decrease in amplitude above $\sim$ a few herz \citep[e.g.][]{Pottschmidt03, Axelsson05, Done07, Kalemci03, Kalemci06}. These negative correlations are often attributed to a low pass filter acting to suppress variability above $\sim$ 2--5 Hz \citep[e.g.][]{Done07, Gierlinski08, Heil11}. On the other hand, at $f \lesssim 2$ Hz, the decrease in the LFQPO amplitude coincides with the growth of the jet, indicating a possible correlation between them. If the jet emits X-rays, considering the decreases in both the LFQPO fractional and absolute amplitudes (Figs \ref{fig:fre2rms}, \ref{fig:rmsxrate}, and \ref{fig:fre2ga2cut}b), the decrease of the LFQPO fractional amplitude might partly be attributed to the increase of the X-ray flux of the jet which is independent of the LFQPO. Even if the jet does not emit X-rays, it might be attributable to the weakening of the LFQPO itself due to some sort of process, for example more accretion material/energy forms the jet but not the corona. \citet{Yan13} also presented a decrease in the LFQPO amplitude that coincides with a possible production of a short-lived jet in GRS~1915+105. Thus our result tends to support the existence of the short-lived jet. Because the low pass filter mainly suppress variability above several herz and the radio flux is inversely correlated with the LFQPO frequency, it might be that both the low pass filter and the jet have impacts on the LFQPO amplitude, and the former plays a dominant role at $f \gtrsim 2$ Hz while the later plays a dominant role at $f \lesssim 2$ Hz. The word ``dominant" here is only for the comparison between the effects of the low pass filter and the jet.

If the LFQPO does come from the corona, then, the negative correlation between the LFQPO frequency and the radio flux indicates a tight correlation between the corona and the jet. In the context of the truncated disc model, this means that decreasing the disc truncation radius leads to a higher QPO frequency \citep[e.g.][]{Done07, Ingram09} and a weaker jet.

For Branch 2, it is interesting to note that the points of the LFQPO absolute amplitude distribute sporadically while the points of the LFQPO amplitude distribute regularly. More intriguingly, the LFQPO amplitude of Branch 2 is roughly in line with that of Branch 1 in the LFQPO amplitude-frequency relationship, hinting that a common mechanism, e.g. a low pass filter, might works.

In summary, we have made a statistical study of both the LFQPO amplitude spectrum and amplitude in GRS~1915+105. The observations are divided into two groups based on the appearance of the LFQPO frequency-HR diagram. The jets of GRS~1915+105 during the two groups of observations are very different. For one group, the LFQPO frequency is negatively correlated with the radio flux. We fitted the LFQPO amplitude spectrum by a power-law with a cutoff, and found that as the LFQPO frequency increases, the spectrum becomes harder. In addition, there is no significant difference between the two groups of observations in the LFQPO amplitude spectrum, indicating that the LFQPO seems not to originate from the jet. The LFQPO amplitude spectrum is hard, suggesting that the LFQPO seems to originate from the corona. As the LFQPO frequency increases, the LFQPO frequency-amplitude relationship evolves from a positive correlation to a negative one, which might be a result of the combined effect of the low pass filter and the jet on the LFQPO amplitude.

\section*{Acknowledgements}
We thank the anonymous referee for comments which greatly improved the quality of the manuscript. We thank Ms. Qian Yang for her help with the English. The research has made use of data obtained from the High Energy Astrophysics Science Archive Research Center (HEASARC), provided by NASA's Goddard Space Flight Center. This work is supported by CAS (KJCX2-YW-T09), 973 Program (2009CB824800), NSFC (11143013, 11173024, 11203063, and 11203064), and WLF of CAS (XBBS201123, LHXZ201201, and XBBS201121).

\bibliographystyle{mn2e}
\bibliography{yan}

\clearpage

\clearpage

\onecolumn
\scriptsize
\begin{longtable}{ccccccccccccc}
\caption{List of GRS 1915+105 Observations suitable for evaluating the LFQPO amplitude spectrum.}
\label{table1}\\
\endfirsthead
\caption{(continued)} \\
\hline
\endhead

\hline\hline
 &  &  &  &  & \multicolumn{3}{c}{LFQPO} & &\multicolumn{3}{c}{LFQPO Amplitude Spectrum$^c$} & \\
\cline{6-8} \cline{10-12} \\
ObsID&Date&GTI$^a$&Count Rate&ChID$^b$&Frequency&Amplitude&$\chi^2$& &$\alpha$&$E_{\rm c}$$^d$&$\chi^2$&Branch$^e$ \\
 &  & (s) & (cts/s/PCU2) &  & (Hz) & (\%rms) &  &  &  & (keV) &  &  \\
\hline
10258-01-02-00 & 29/07/96& 9160&1739&Ch1E3& $0.697\pm0.002$& $10.8\pm0.4$& 2.01& &$-0.39\pm0.06$&    No cutoff & 1.05& B$_1$\\
10258-01-03-00a& 06/08/96& 3328&1757&Ch1E3& $1.687\pm0.005$& $12.5\pm0.5$& 2.65& &$-0.51\pm0.03$& $49.6\pm 5.6$& 0.17& B$_1$\\
10258-01-03-00b& 06/08/96& 3360&1771&Ch1E3& $1.332\pm0.003$& $12.4\pm0.7$& 2.18& &$-0.39\pm0.03$& $81.3\pm17.2$& 0.15& B$_1$\\
10258-01-03-00c& 06/08/96& 3360&1736&Ch1E3& $1.453\pm0.003$& $12.5\pm0.5$& 2.25& &$-0.50\pm0.07$& $54.9\pm16.5$& 0.77& B$_1$\\
10258-01-04-00a& 14/08/96& 6800&1915&Ch1E3& $2.694\pm0.003$& $12.7\pm0.2$& 3.13& &$-0.58\pm0.04$& $45.3\pm 7.8$& 1.44& B$_1$\\
10258-01-04-00b& 14/08/96& 3408&1971&Ch1E3& $3.133\pm0.007$& $12.1\pm0.3$& 2.18& &$-0.61\pm0.06$& $48.8\pm13.2$& 1.41& B$_1$\\
10258-01-05-00a& 20/08/96& 2688&3743&Ch2E3& $6.370\pm0.030$&  $5.4\pm0.2$& 1.74& &$-1.19\pm0.22$& $18.4\pm 6.1$& 1.29& B$_1$\\
10258-01-05-00b& 20/08/96& 3376&3750&Ch2E3& $6.359\pm0.024$&  $5.3\pm0.2$& 2.22& &$-1.20\pm0.20$& $19.5\pm 5.6$& 0.64& B$_1$\\
10258-01-06-00a& 29/08/96& 1400&5549&Ch2E3& $7.338\pm0.038$&  $2.4\pm0.2$& 1.22& &$-0.84\pm0.26$&    No cutoff & 2.00& B$_1$\\
10258-01-06-00b& 29/08/96& 3408&5587&Ch2E3& $7.560\pm0.024$&  $2.2\pm0.1$& 1.80& &$-1.08\pm0.28$& $32.3\pm23.1$& 2.77& B$_1$\\
10408-01-22-00 & 11/07/96& 3328&2122&Ch2E3& $3.476\pm0.005$& $10.2\pm0.3$& 1.03& &$-0.78\pm0.05$& $30.1\pm 3.5$& 0.49& B$_1$\\
10408-01-22-01 & 11/07/96& 3312&2020&Ch2E3& $2.780\pm0.005$& $11.9\pm0.3$& 1.61& &$-0.69\pm0.06$& $34.0\pm 5.4$& 0.96& B$_1$\\
10408-01-22-02a& 11/07/96& 1600&1989&Ch2E3& $2.547\pm0.008$& $12.0\pm0.5$& 1.67& &$-0.74\pm0.04$& $31.7\pm 3.2$& 0.20& B$_1$\\
10408-01-22-02b& 11/07/96&  820&1954&Ch2E3& $2.509\pm0.008$& $12.0\pm0.8$& 1.79& &$-0.89\pm0.02$& $22.6\pm 1.0$& 0.03& B$_1$\\
10408-01-22-02c& 11/07/96&  892&1929&Ch2E3& $2.623\pm0.009$& $11.8\pm0.6$& 1.56& &$-0.74\pm0.03$& $33.0\pm 2.2$& 0.05& B$_1$\\
10408-01-23-00a& 14/07/96& 3167&2109&Ch2E3& $3.501\pm0.006$& $11.2\pm0.3$& 1.66& &$-0.71\pm0.02$& $33.2\pm 1.8$& 0.14& B$_1$\\
10408-01-23-00b& 14/07/96& 3312&2108&Ch2E3& $3.611\pm0.005$& $11.0\pm0.3$& 2.02& &$-0.74\pm0.02$& $31.9\pm 2.0$& 0.19& B$_1$\\
10408-01-23-00c& 14/07/96& 3257&2255&Ch2E3& $4.178\pm0.008$& $10.3\pm0.3$& 1.51& &$-0.69\pm0.03$& $36.7\pm 3.8$& 0.34& B$_1$\\
10408-01-24-00a& 16/07/96& 2447&1949&Ch2E3& $2.242\pm0.006$& $12.7\pm0.5$& 2.88& &$-0.58\pm0.02$& $45.5\pm 3.9$& 0.11& B$_1$\\
10408-01-24-00b& 16/07/96& 3312&1943&Ch2E3& $2.324\pm0.005$& $13.2\pm0.4$& 2.44& &$-0.69\pm0.02$& $30.6\pm 1.5$& 0.11& B$_1$\\
10408-01-24-00c& 16/07/96& 2953&1952&Ch2E3& $2.541\pm0.004$& $12.2\pm0.4$& 2.61& &$-0.59\pm0.06$& $44.5\pm 9.3$& 0.75& B$_1$\\
10408-01-24-00d& 16/07/96&  913&1965&Ch2E3& $2.597\pm0.007$& $12.0\pm0.7$& 1.65& &$-0.61\pm0.05$& $44.2\pm 7.6$& 0.19& B$_1$\\
10408-01-25-00 & 19/07/96& 9952&1820&Ch1E3& $1.130\pm0.002$& $12.7\pm0.3$& 2.46& &$-0.47\pm0.04$& $65.0\pm14.1$& 0.80& B$_1$\\
10408-01-27-00a& 26/07/96& 2336&1783&Ch1E3& $0.645\pm0.002$& $10.6\pm0.9$& 1.42& &$-0.55\pm0.11$& $63.2\pm26.3$& 0.15& B$_1$\\
10408-01-27-00b& 26/07/96& 3296&1791&Ch1E3& $0.618\pm0.002$&  $9.3\pm0.7$& 1.07& &$-0.53\pm0.05$& $57.0\pm11.7$& 0.15& B$_1$\\
10408-01-27-00c& 26/07/96& 3296&1769&Ch1E3& $0.629\pm0.003$&  $9.7\pm0.6$& 1.28& &$-0.41\pm0.04$& $79.5\pm19.4$& 0.11& B$_1$\\
10408-01-28-00a& 03/08/96& 3328&1742&Ch1E3& $0.996\pm0.002$& $11.8\pm0.6$& 1.61& &$-0.40\pm0.05$& $93.8\pm36.2$& 0.30& B$_1$\\
10408-01-28-00b& 03/08/96& 3328&1744&Ch1E3& $0.964\pm0.004$& $11.2\pm0.6$& 1.13& &$-0.36\pm0.05$&    No cutoff & 0.28& B$_1$\\
10408-01-28-00c& 03/08/96& 3328&1731&Ch1E3& $0.926\pm0.002$& $12.2\pm0.6$& 1.49& &$-0.34\pm0.05$&    No cutoff & 0.37& B$_1$\\
10408-01-29-00a& 10/08/96& 2965&1760&Ch1E3& $1.664\pm0.003$& $12.3\pm0.5$& 1.64& &$-0.55\pm0.05$& $51.9\pm11.0$& 0.43& B$_1$\\
10408-01-29-00b& 10/08/96& 3392&1784&Ch1E3& $1.857\pm0.004$& $12.3\pm0.6$& 1.73& &$-0.57\pm0.05$& $65.0\pm15.4$& 0.37& B$_1$\\
10408-01-29-00c& 10/08/96& 3392&1787&Ch1E3& $1.954\pm0.004$& $12.4\pm0.5$& 1.56& &$-0.53\pm0.07$& $52.4\pm16.2$& 0.93& B$_1$\\
10408-01-30-00a& 18/08/96& 1696&2388&Ch1E3& $4.316\pm0.013$&  $9.3\pm0.3$& 1.64& &$-0.82\pm0.04$& $29.2\pm 3.0$& 0.23& B$_1$\\
10408-01-30-00b& 18/08/96& 1696&2588&Ch1E3& $4.794\pm0.012$&  $8.1\pm0.3$& 1.29& &$-0.71\pm0.03$& $42.4\pm 6.4$& 0.14& B$_1$\\
10408-01-30-00c& 18/08/96& 1696&2842&Ch1E3& $5.204\pm0.017$&  $7.1\pm0.3$& 1.40& &$-0.86\pm0.10$& $23.8\pm 6.2$& 0.83& B$_1$\\
10408-01-30-00d& 18/08/96& 1696&2752&Ch1E3& $4.902\pm0.012$&  $7.4\pm0.4$& 1.07& &$-0.79\pm0.04$& $31.4\pm 3.6$& 0.15& B$_1$\\
10408-01-30-00e& 18/08/96& 1688&2986&Ch1E3& $5.431\pm0.014$&  $5.5\pm0.3$& 1.15& &$-0.64\pm0.07$& $54.6\pm20.8$& 0.36& B$_1$\\
10408-01-31-00a& 25/08/96& 2319&2327&Ch1E3& $4.101\pm0.006$&  $9.5\pm0.3$& 1.72& &$-0.76\pm0.08$& $36.2\pm 9.8$& 1.33& B$_1$\\
10408-01-31-00b& 25/08/96& 1000&2555&Ch1E3& $4.672\pm0.014$&  $8.0\pm0.4$& 1.32& &$-0.87\pm0.09$& $21.6\pm 4.4$& 0.55& B$_1$\\
10408-01-31-00c& 25/08/96& 1328&2496&Ch1E3& $4.487\pm0.014$&  $8.5\pm0.3$& 1.49& &$-0.94\pm0.08$& $18.7\pm 2.9$& 0.77& B$_1$\\
10408-01-31-00d& 25/08/96& 1000&2323&Ch1E3& $4.172\pm0.012$&  $9.4\pm0.4$& 2.06& &$-0.82\pm0.05$& $30.6\pm 4.2$& 0.29& B$_1$\\
10408-01-31-00e& 25/08/96& 1664&2133&Ch1E3& $3.632\pm0.008$& $10.3\pm0.4$& 1.39& &$-0.75\pm0.06$& $36.2\pm 6.5$& 0.51& B$_1$\\
10408-01-31-00f& 25/08/96& 1664&2057&Ch1E3& $3.388\pm0.007$& $10.8\pm0.4$& 1.69& &$-0.82\pm0.05$& $26.4\pm 3.0$& 0.31& B$_1$\\
10408-01-32-00a& 31/08/96& 2912&4239&Ch1E3& $6.654\pm0.033$&  $3.2\pm0.2$& 2.25& &$-1.25\pm0.33$& $24.2\pm17.7$& 6.35& B$_1$\\
10408-01-32-00b& 31/08/96& 3312&3648&Ch1E3& $5.965\pm0.019$&  $4.4\pm0.2$& 2.38& &$-1.09\pm0.15$& $20.4\pm 6.7$& 1.82& B$_1$\\
10408-01-32-00c& 31/08/96& 1170&3314&Ch1E3& $5.674\pm0.029$&  $5.4\pm0.3$& 1.60& &$-0.93\pm0.03$& $30.4\pm 3.7$& 0.02& B$_1$\\
10408-01-33-00a& 07/09/96&  912&3527&Ch1E3& $5.610\pm0.034$&  $5.3\pm0.3$& 1.67& &$-0.63\pm0.12$&    No cutoff & 0.52& B$_1$\\
10408-01-33-00b& 07/09/96& 2495&3743&Ch1E3& $5.708\pm0.022$&  $4.3\pm0.2$& 1.79& &$-1.04\pm0.19$& $25.2\pm11.7$& 2.43& B$_1$\\
10408-01-33-00c& 07/09/96& 1295&3655&Ch1E3& $5.542\pm0.040$&  $5.0\pm0.3$& 1.91& &$-0.77\pm0.09$&    No cutoff & 2.75& B$_1$\\
10408-01-42-00a& 23/10/96& 3312&3289&Ch1E3& $5.063\pm0.010$&  $5.6\pm0.2$& 2.44& &$-0.80\pm0.09$& $23.7\pm 5.1$& 1.09& B$_2$\\
10408-01-42-00b& 23/10/96& 3312&2921&Ch1E3& $4.709\pm0.010$&  $6.6\pm0.2$& 2.20& &$-0.57\pm0.08$&    No cutoff & 0.76& B$_2$\\
10408-01-43-00a& 23/10/96& 2416&3274&Ch1E3& $5.020\pm0.011$&  $5.7\pm0.2$& 1.85& &$-1.02\pm0.10$& $17.3\pm 3.3$& 0.56& B$_2$\\
10408-01-43-00b& 23/10/96& 2284&3314&Ch1E3& $5.077\pm0.013$&  $5.6\pm0.2$& 2.01& &$-0.60\pm0.20$&    No cutoff & 2.84& B$_2$\\
10408-01-43-00c& 23/10/96& 1980&3302&Ch1E3& $5.135\pm0.013$&  $5.5\pm0.2$& 1.67& &$-0.70\pm0.12$& $27.6\pm 9.6$& 0.89& B$_2$\\
10408-01-43-00d& 23/10/96& 1740&2709&Ch1E3& $4.462\pm0.014$&  $7.2\pm0.3$& 1.91& &$-0.74\pm0.12$& $33.1\pm12.9$& 1.02& B$_2$\\
20186-03-02-052a&17/09/97& 3031&3096&Ch3E3& $5.390\pm0.016$&  $5.3\pm0.3$& 1.56& &$-1.12\pm0.07$& $17.1\pm 2.1$& 0.38& B$_1$\\
20186-03-02-052b&17/09/97& 3031&3203&Ch3E3& $5.818\pm0.019$&  $5.9\pm0.2$& 1.95& &$-1.05\pm0.14$& $25.5\pm11.0$& 1.20& B$_1$\\
20186-03-02-052c&17/09/97& 3312&2348&Ch3E3& $4.086\pm0.006$&  $9.3\pm0.2$& 1.34& &$-1.00\pm0.06$& $21.4\pm 2.5$& 1.46& B$_1$\\
20186-03-02-052d&17/09/97& 3312&2563&Ch3E3& $4.634\pm0.008$&  $8.2\pm0.2$& 1.16& &$-1.06\pm0.06$& $18.6\pm 2.2$& 1.29& B$_1$\\
20186-03-02-052e&17/09/97& 3312&2802&Ch3E3& $5.157\pm0.011$&  $6.5\pm0.2$& 1.43& &$-1.01\pm0.06$& $20.5\pm 2.8$& 0.81& B$_1$\\
20186-03-02-060a&18/09/97& 2768&2852&Ch3E3& $4.984\pm0.022$&  $7.6\pm0.3$& 1.00& &$-0.81\pm0.08$& $42.9\pm16.8$& 0.75& B$_1$\\
20186-03-02-060b&18/09/97& 9936&4385&Ch3E3& $6.761\pm0.036$&  $4.7\pm0.1$& 1.55& &$-0.71\pm0.03$&    No cutoff & 1.34& B$_1$\\
20186-03-02-060c&18/09/97& 3312&2679&Ch3E3& $4.788\pm0.011$&  $7.9\pm0.2$& 1.33& &$-0.87\pm0.05$& $31.7\pm 5.3$& 0.62& B$_1$\\
20186-03-02-06a& 18/09/97& 1656&2767&Ch3E3& $5.042\pm0.017$&  $7.3\pm0.2$& 1.17& &$-0.97\pm0.12$& $24.0\pm 6.5$& 1.66& B$_1$\\
20186-03-02-06b& 18/09/97& 1656&2430&Ch3E3& $4.240\pm0.012$&  $8.5\pm0.3$& 1.50& &$-0.86\pm0.05$& $28.4\pm 4.3$& 0.67& B$_1$\\
20186-03-02-06c& 18/09/97& 1600&2255&Ch3E3& $3.820\pm0.009$& $10.0\pm0.4$& 1.42& &$-1.02\pm0.04$& $18.9\pm 1.4$& 0.40& B$_1$\\
20186-03-02-06d& 18/09/97& 1695&2399&Ch3E3& $4.291\pm0.009$&  $9.0\pm0.3$& 1.38& &$-1.07\pm0.07$& $18.5\pm 2.6$& 0.86& B$_1$\\
20186-03-02-06e& 18/09/97& 1550&2761&Ch3E3& $5.060\pm0.014$&  $7.0\pm0.3$& 1.28& &$-1.01\pm0.11$& $21.7\pm 5.6$& 1.06& B$_1$\\
20186-03-02-06f& 18/09/97& 1569&3648&Ch3E3& $5.949\pm0.060$&  $6.1\pm0.3$& 1.32& &$-0.72\pm0.13$&    No cutoff & 0.79& B$_1$\\
20402-01-05-00 & 05/12/96& 2048&1421&Ch5E3& $2.819\pm0.004$& $13.2\pm0.3$& 1.38& &$-0.82\pm0.10$& $21.9\pm 4.3$& 8.10& B$_2$\\
20402-01-06-00a& 11/12/96& 3312&1360&Ch5E3& $3.032\pm0.009$& $12.7\pm0.5$& 1.13& &$-0.87\pm0.10$& $19.8\pm 3.7$& 2.68& B$_2$\\
20402-01-06-00b& 11/12/96& 3312&1279&Ch5E3& $2.837\pm0.007$& $13.2\pm0.5$& 1.23& &$-0.74\pm0.07$& $25.2\pm 4.0$& 2.03& B$_2$\\
20402-01-06-00c& 11/12/96& 2780&1211&Ch5E3& $2.569\pm0.007$& $13.0\pm0.5$& 1.40& &$-0.78\pm0.09$& $25.0\pm 4.8$& 2.46& B$_2$\\
20402-01-07-00 & 19/12/96& 9296&1310&Ch5E3& $3.116\pm0.005$& $13.0\pm0.2$& 1.78& &$-0.89\pm0.07$& $17.4\pm 2.4$& 4.31& B$_2$\\
20402-01-08-00a& 24/12/96& 2658&1318&Ch5E3& $3.859\pm0.010$& $10.7\pm0.3$& 1.52& &$-0.81\pm0.09$& $20.7\pm 4.5$& 1.22& B$_2$\\
20402-01-08-00b& 24/12/96& 2834&1325&Ch5E3& $3.934\pm0.010$& $10.6\pm0.3$& 2.26& &$-0.87\pm0.08$& $18.6\pm 3.3$& 1.23& B$_2$\\
20402-01-08-01 & 25/12/96& 3312&1232&Ch5E3& $3.469\pm0.009$& $12.0\pm0.3$& 1.35& &$-0.72\pm0.10$& $28.6\pm 8.3$& 2.45& B$_2$\\
20402-01-09-00 & 31/12/96& 7548&1099&Ch5E3& $2.816\pm0.006$& $12.9\pm0.3$& 1.77& &$-0.73\pm0.08$& $23.4\pm 4.4$& 4.50& B$_2$\\
20402-01-10-00 & 08/01/97& 9804& 993&Ch5E3& $2.912\pm0.006$& $12.6\pm0.2$& 2.07& &$-0.77\pm0.08$& $22.2\pm 3.7$& 5.10& B$_2$\\
20402-01-11-00 & 14/01/97& 6519& 912&Ch5E3& $2.919\pm0.007$& $11.7\pm0.3$& 1.53& &$-0.84\pm0.11$& $19.7\pm 4.0$& 4.46& B$_2$\\
20402-01-12-00a& 23/01/97& 5695& 883&Ch5E3& $2.802\pm0.006$& $12.1\pm0.4$& 1.37& &$-0.76\pm0.07$& $23.0\pm 3.2$& 1.34& B$_2$\\
20402-01-12-00b& 23/01/97& 3755& 894&Ch5E3& $2.783\pm0.007$& $11.7\pm0.5$& 1.50& &$-0.75\pm0.11$& $23.9\pm 6.2$& 1.83& B$_2$\\
20402-01-13-00 & 29/01/97&10000& 936&Ch5E3& $3.650\pm0.007$& $11.7\pm0.2$& 2.02& &$-0.82\pm0.15$& $20.5\pm 6.5$& 13.4& B$_2$\\
20402-01-14-00 & 01/02/97& 9394& 910&Ch5E3& $3.566\pm0.007$& $11.6\pm0.2$& 2.00& &$-0.82\pm0.10$& $19.9\pm 4.1$& 5.31& B$_2$\\
20402-01-15-00 & 09/02/97&10222& 816&Ch5E3& $2.260\pm0.004$& $12.2\pm0.3$& 1.71& &$-0.55\pm0.08$& $37.3\pm 9.8$& 4.15& B$_2$\\
20402-01-16-00 & 22/02/97& 5951& 803&Ch5E3& $2.977\pm0.007$& $11.1\pm0.3$& 1.21& &$-0.62\pm0.06$& $28.4\pm 4.8$& 1.31& B$_2$\\
20402-01-20-00 & 17/03/97& 7300& 807&Ch5E3& $3.208\pm0.006$& $11.1\pm0.3$& 1.43& &$-0.75\pm0.13$& $24.1\pm 7.5$& 5.50& B$_2$\\
20402-01-26-00a& 25/04/97& 2220&1137&Ch5E3& $3.959\pm0.012$& $10.6\pm0.3$& 1.76& &$-0.92\pm0.10$& $16.2\pm 2.7$& 1.17& B$_2$\\
20402-01-26-00b& 25/04/97& 2884&1188&Ch5E3& $4.286\pm0.010$& $10.3\pm0.3$& 2.08& &$-0.97\pm0.10$& $15.5\pm 2.7$& 1.07& B$_2$\\
20402-01-26-00c& 25/04/97& 3300&1210&Ch5E3& $4.468\pm0.016$& $10.0\pm0.3$& 1.94& &$-0.93\pm0.14$& $21.0\pm 6.1$& 2.30& B$_2$\\
20402-01-26-00d& 25/04/97& 3328&1178&Ch5E3& $4.258\pm0.017$&  $9.7\pm0.3$& 1.74& &$-0.69\pm0.14$& $41.3\pm23.3$& 2.46& B$_2$\\
20402-01-26-00e& 25/04/97& 1964&1163&Ch5E3& $4.391\pm0.014$&  $9.9\pm0.3$& 2.08& &$-0.83\pm0.10$& $21.4\pm 4.8$& 1.04& B$_2$\\
20402-01-48-00a& 29/09/97& 3296&4714&Ch5E3& $7.589\pm0.036$&  $2.6\pm0.1$& 1.56& &$-0.96\pm0.08$&    No cutoff & 1.39& B$_1$\\
20402-01-48-00b& 29/09/97& 3328&2726&Ch5E3& $4.712\pm0.014$&  $6.7\pm0.3$& 1.60& &$-0.95\pm0.07$& $24.5\pm 4.6$& 0.60& B$_1$\\
20402-01-50-01 & 16/10/97& 4994&1497&Ch5E3& $1.047\pm0.003$& $11.0\pm0.5$& 2.16& &$-0.57\pm0.03$& $45.6\pm 4.8$& 0.25& B$_1$\\
20402-01-51-00 & 22/10/97& 9399&1490&Ch5E3& $1.396\pm0.002$& $12.6\pm0.3$& 3.30& &$-0.57\pm0.03$& $50.4\pm 5.7$& 0.73& B$_1$\\
30182-01-01-00 & 08/07/98&11606&1435&Ch3E3& $2.139\pm0.003$& $15.6\pm0.3$& 2.09& &$-0.73\pm0.06$& $31.5\pm 5.2$& 5.23& B$_2$\\
30182-01-02-00a& 09/07/98& 5073&1889&Ch3E3& $3.248\pm0.005$& $12.7\pm0.3$& 1.57& &$-0.89\pm0.08$& $22.0\pm 3.7$& 4.96& B$_2$\\
30182-01-02-00b& 09/07/98& 3359&2069&Ch3E3& $3.544\pm0.006$& $11.9\pm0.3$& 1.55& &$-0.89\pm0.09$& $21.7\pm 3.9$& 4.17& B$_2$\\
30182-01-02-00c& 09/07/98& 2968&2466&Ch3E3& $3.975\pm0.008$&  $9.9\pm0.3$& 1.73& &$-0.90\pm0.07$& $21.0\pm 2.9$& 1.48& B$_2$\\
30182-01-03-00a& 10/07/98& 3344&3479&Ch3E3& $5.097\pm0.014$&  $5.7\pm0.2$& 2.54& &$-0.93\pm0.06$& $25.3\pm 3.6$& 0.36& B$_2$\\
30182-01-03-00b& 10/07/98& 2472&3677&Ch3E3& $5.166\pm0.011$&  $5.0\pm0.2$& 1.91& &$-0.78\pm0.07$& $29.1\pm 6.4$& 0.45& B$_2$\\
30182-01-04-00a& 11/07/98& 1678&2360&Ch3E3& $4.110\pm0.010$&  $9.2\pm0.3$& 1.54& &$-1.04\pm0.10$& $18.3\pm 3.4$& 1.93& B$_2$\\
30182-01-04-00b& 11/07/98& 4166&1933&Ch3E3& $3.403\pm0.006$& $11.5\pm0.3$& 1.65& &$-1.03\pm0.08$& $18.5\pm 2.5$& 3.99& B$_2$\\
30182-01-04-00c& 11/07/98& 3328&1709&Ch3E3& $2.918\pm0.005$& $12.3\pm0.4$& 1.86& &$-0.97\pm0.08$& $20.5\pm 3.1$& 3.39& B$_2$\\
30182-01-04-00d& 11/07/98& 3324&1604&Ch3E3& $2.665\pm0.005$& $13.3\pm0.4$& 1.63& &$-0.89\pm0.09$& $23.4\pm 4.3$& 3.77& B$_2$\\
30182-01-04-01a& 12/07/98& 2236&1581&Ch3E3& $2.641\pm0.006$& $13.6\pm0.7$& 1.03& &$-0.87\pm0.09$& $21.8\pm 4.1$& 2.26& B$_2$\\
30182-01-04-01b& 12/07/98& 2728&1513&Ch3E3& $2.411\pm0.005$& $14.4\pm0.5$& 1.45& &$-0.83\pm0.08$& $26.0\pm 4.7$& 2.31& B$_2$\\
30182-01-04-01c& 12/07/98& 3340&1605&Ch3E3& $2.723\pm0.006$& $13.9\pm0.4$& 1.18& &$-0.83\pm0.08$& $26.2\pm 4.8$& 3.31& B$_2$\\
30182-01-04-01d& 12/07/98& 3340&2010&Ch3E3& $3.377\pm0.013$& $13.5\pm0.3$& 1.44& &$-0.89\pm0.11$& $20.4\pm 4.2$& 4.72& B$_2$\\
30182-01-04-01e& 12/07/98& 2400&2665&Ch3E3& $4.222\pm0.010$&  $8.5\pm0.3$& 2.13& &$-0.85\pm0.10$& $21.9\pm 4.7$& 1.94& B$_2$\\
30402-01-09-01 & 10/04/98& 2546&1979&Ch6E3& $2.157\pm0.004$& $12.9\pm0.4$& 2.44& &$-0.68\pm0.04$& $31.0\pm 3.7$& 0.72& B$_1$\\
30402-01-10-00a& 11/04/98& 3312&1970&Ch6E3& $1.590\pm0.003$& $12.8\pm0.6$& 1.14& &$-0.60\pm0.04$& $41.3\pm 6.4$& 0.60& B$_1$\\
30402-01-10-00b& 11/04/98& 6303&1956&Ch6E3& $1.722\pm0.003$& $12.7\pm0.3$& 3.77& &$-0.60\pm0.03$& $40.3\pm 4.6$& 0.79& B$_1$\\
30402-01-11-00a& 20/04/98& 3311&2777&Ch6E3& $5.401\pm0.013$&  $7.5\pm0.2$& 2.42& &$-0.87\pm0.09$& $22.5\pm 4.8$& 1.15& B$_1$\\
30402-01-11-00b& 20/04/98& 2271&2952&Ch6E3& $5.827\pm0.018$&  $6.9\pm0.2$& 1.74& &$-0.80\pm0.13$& $26.6\pm10.2$& 1.57& B$_1$\\
30703-01-16-00 & 28/04/98& 5038&1816&Ch6E3& $1.376\pm0.003$& $12.6\pm0.4$& 1.40& &$-0.65\pm0.05$& $33.1\pm 4.9$& 1.41& B$_1$\\
30703-01-17-00 & 06/05/98& 4584&1739&Ch6E3& $0.926\pm0.002$& $11.8\pm0.5$& 1.67& &$-0.51\pm0.05$& $48.9\pm 9.3$& 0.68& B$_1$\\
30703-01-22-00 & 27/06/98& 3375&1539&Ch6E3& $2.253\pm0.005$& $14.2\pm0.5$& 1.98& &$-0.73\pm0.03$& $31.7\pm 3.1$& 0.60& B$_2$\\
30703-01-25-00a& 23/07/98& 2626&1718&Ch6E3& $3.175\pm0.007$& $12.6\pm0.4$& 1.21& &$-0.90\pm0.11$& $20.9\pm 4.2$& 4.66& B$_2$\\
30703-01-25-00b& 23/07/98& 2322&2146&Ch6E3& $3.810\pm0.009$& $10.3\pm0.3$& 1.31& &$-0.90\pm0.10$& $18.5\pm 3.5$& 2.36& B$_2$\\
30703-01-33-00 & 15/09/98& 4917&1400&Ch6E3& $3.297\pm0.007$& $12.5\pm0.3$& 1.21& &$-0.82\pm0.09$& $20.3\pm 3.5$& 4.40& B$_2$\\
30703-01-41-00 & 26/12/98& 4707&1233&Ch6E3& $2.154\pm0.004$& $14.8\pm0.4$& 1.59& &$-0.59\pm0.05$& $38.9\pm 7.0$& 1.69& B$_2$\\
40403-01-08-00 & 02/06/99& 9884&1584&Ch6E4& $2.475\pm0.003$& $13.6\pm0.3$& 3.25& &$-0.72\pm0.06$& $34.3\pm 5.5$& 4.39& B$_2$\\
40403-01-09-00 & 08/07/99&13355&1343&Ch6E4& $2.030\pm0.003$& $15.6\pm0.3$& 2.08& &$-0.66\pm0.08$& $39.9\pm10.0$& 9.89& B$_2$\\
40403-01-11-00 & 28/02/00&13355&2426&Ch6E4& $4.339\pm0.010$&  $8.0\pm0.3$& 2.38& &$-0.68\pm0.09$& $34.5\pm10.9$& 1.41& B$_2$\\
40703-01-01-00 & 01/01/99& 9731&1281&Ch6E4& $2.264\pm0.003$& $15.0\pm0.3$& 1.88& &$-0.69\pm0.07$& $29.2\pm 5.1$& 4.92& B$_2$\\
40703-01-02-00 & 08/01/99& 9005&1861&Ch6E4& $3.568\pm0.005$& $11.5\pm0.2$& 1.80& &$-0.72\pm0.08$& $27.8\pm 6.3$& 7.23& B$_2$\\
40703-01-05-00 & 12/02/99&10129&1592&Ch6E4& $4.204\pm0.006$& $10.0\pm0.1$& 2.51& &$-0.85\pm0.09$& $19.7\pm 3.5$& 5.77& B$_2$\\
40703-01-09-00 & 28/03/99& 4702&1418&Ch6E4& $2.782\pm0.005$& $12.9\pm0.3$& 1.11& &$-0.76\pm0.08$& $24.9\pm 4.2$& 2.67& B$_2$\\
40703-01-38-02 & 15/11/99& 2501&5138&Ch6E4& $7.978\pm0.036$&  $2.9\pm0.1$& 1.12& &$-0.94\pm0.04$&    No cutoff & 0.45& B$_1$\\
50125-01-01-03 & 13/07/00& 2735&1747&Ch3E5& $3.021\pm0.006$& $13.0\pm0.4$& 1.62& &$-0.73\pm0.07$& $35.0\pm 7.2$& 2.29& B$_2$\\
50125-01-03-00a& 15/07/00& 4348&2077&Ch3E5& $3.548\pm0.007$& $12.1\pm0.3$& 1.64& &$-0.80\pm0.07$& $27.3\pm 4.6$& 3.26& B$_2$\\
50125-01-03-00b& 15/07/00&10652&1818&Ch3E5& $3.184\pm0.004$& $12.9\pm0.2$& 3.25& &$-0.81\pm0.07$& $26.7\pm 4.2$& 7.47& B$_2$\\
50703-01-01-00 & 08/03/00& 4755&1314&Ch6E4& $2.343\pm0.007$& $15.9\pm0.5$& 1.22& &$-0.61\pm0.07$& $31.6\pm 6.4$& 2.15& B$_2$\\
50703-01-49-00 & 27/02/01& 5467&1434&Ch6E5& $2.611\pm0.004$& $13.2\pm0.3$& 2.19& &$-0.85\pm0.10$& $27.0\pm 6.2$& 5.46& B$_2$\\
50703-01-55-01 & 17/04/01& 6896&1583&Ch6E5& $2.839\pm0.004$& $13.8\pm0.3$& 2.00& &$-0.70\pm0.09$& $34.5\pm 9.3$& 8.04& B$_2$\\
50703-01-67-00 & 22/07/01& 1806&1243&Ch6E5& $2.183\pm0.005$& $14.6\pm0.8$& 1.33& &$-0.65\pm0.07$& $41.7\pm 9.9$& 1.16& B$_2$\\
60100-01-01-00 & 05/08/01& 3280&1249&Ch6E5& $2.225\pm0.004$& $14.3\pm0.6$& 1.42& &$-0.60\pm0.09$& $50.6\pm18.7$& 2.63& B$_2$\\
60100-01-02-000a&06/08/01& 2748&1487&Ch6E5& $2.712\pm0.005$& $14.6\pm0.5$& 1.29& &$-0.60\pm0.08$& $34.0\pm 8.6$& 2.86& B$_2$\\
60100-01-02-000b&06/08/01& 2496&1654&Ch6E5& $3.017\pm0.006$& $13.5\pm0.5$& 1.11& &$-0.58\pm0.11$& $40.2\pm16.0$& 3.84& B$_2$\\
60100-01-02-000c&06/08/01& 2648&1762&Ch6E5& $3.201\pm0.007$& $12.9\pm0.4$& 1.47& &$-0.75\pm0.10$& $23.2\pm 5.3$& 3.36& B$_2$\\
60100-01-02-000d&06/08/01& 2816&1967&Ch6E5& $3.516\pm0.007$& $11.3\pm0.3$& 1.38& &$-0.61\pm0.11$& $43.9\pm21.5$& 4.36& B$_2$\\
60100-01-02-000e&06/08/01& 2964&2178&Ch6E5& $3.845\pm0.009$& $10.1\pm0.3$& 2.28& &$-0.72\pm0.06$& $27.2\pm 4.7$& 1.10& B$_2$\\
60405-01-03-00 & 05/08/01& 6560&1474&Ch6E5& $2.729\pm0.004$& $14.3\pm0.3$& 1.56& &$-0.66\pm0.09$& $29.8\pm 7.3$& 5.78& B$_2$\\
60701-01-16-00 & 28/02/02& 3068&1820&Ch6E5& $0.377\pm0.002$&  $7.6\pm0.6$& 0.73& &$-0.35\pm0.05$&    No cutoff & 0.17& B$_1$\\
60701-01-16-01 & 28/02/02& 3109&1809&Ch6E5& $0.395\pm0.002$&  $8.3\pm0.9$& 1.02& &$-0.45\pm0.11$& $50.2\pm29.3$& 0.21& B$_1$\\
60701-01-23-00 & 22/01/02& 3263&1986&Ch6E5& $2.073\pm0.003$& $13.0\pm0.4$& 2.92& &$-0.82\pm0.05$& $21.7\pm 2.5$& 0.42& B$_1$\\
60701-01-28-00 & 06/03/02& 9680&1744&Ch6E5& $0.466\pm0.001$&  $9.0\pm0.5$& 1.49& &$-0.31\pm0.04$&    No cutoff & 0.40& B$_1$\\
60701-01-33-00 & 24/04/02& 3247&1426&Ch6E5& $1.029\pm0.002$& $10.8\pm0.6$& 1.09& &$-0.46\pm0.07$& $99.0\pm59.0$& 0.91& B$_1$\\
70702-01-23-00 & 03/10/02& 3231&1931&Ch6E5& $3.449\pm0.005$& $12.1\pm0.4$& 1.54& &$-0.69\pm0.13$& $28.3\pm 9.5$& 3.98& B$_2$\\
70702-01-24-00 & 09/10/02& 3264&1328&Ch6E5& $2.581\pm0.005$& $12.9\pm0.5$& 2.51& &$-0.74\pm0.10$& $41.6\pm13.8$& 2.86& B$_2$\\
70703-01-01-08 & 01/04/02&10704&1902&Ch3E5& $2.589\pm0.004$& $11.7\pm0.2$& 4.27& &$-0.69\pm0.07$& $34.1\pm 7.3$& 3.56& B$_1$\\
70703-01-01-14 & 29/03/02& 8240&1869&Ch6E5& $2.634\pm0.004$& $12.6\pm0.2$& 4.02& &$-0.65\pm0.03$& $37.3\pm 4.4$& 1.21& B$_1$\\
80127-02-03-00 & 10/04/03&11728&1884&Ch4E5& $1.088\pm0.002$& $13.3\pm0.4$& 2.18& &$-0.40\pm0.05$& $97.9\pm38.1$& 2.64& B$_1$\\
80701-01-08-00 & 25/10/06& 3216&2375&Ch6E5& $4.652\pm0.012$&  $8.0\pm0.2$& 1.51& &$-0.97\pm0.13$& $20.2\pm 5.5$& 1.57& B$_2$\\
80701-01-26-00 & 28/11/06& 6304&1334&Ch6E5& $2.543\pm0.003$& $14.2\pm0.4$& 2.24& &$-0.75\pm0.08$& $25.2\pm 4.8$& 1.99& B$_2$\\
80701-01-32-00 & 04/12/06& 6239&1212&Ch6E5& $2.102\pm0.003$& $15.6\pm0.5$& 1.75& &$-0.49\pm0.09$& $41.7\pm12.6$& 5.33& B$_2$\\
80701-01-51-00 & 09/12/06& 6960&1252&Ch6E5& $2.222\pm0.003$& $15.0\pm0.4$& 1.91& &$-0.58\pm0.07$& $39.0\pm 9.7$& 2.99& B$_2$\\
80701-01-55-02 & 11/01/07& 5440&1131&Ch6E5& $2.609\pm0.003$& $14.4\pm0.4$& 1.87& &$-0.60\pm0.10$& $30.0\pm 8.3$& 6.07& B$_2$\\
80701-01-56-00 & 18/01/07& 9600&1073&Ch6E5& $2.557\pm0.002$& $13.2\pm0.3$& 3.58& &$-0.68\pm0.08$& $47.6\pm14.5$& 4.03& B$_2$\\
80701-01-57-00 & 24/01/07& 9584&1100&Ch6E5& $2.063\pm0.004$& $13.6\pm0.4$& 3.49& &$-0.43\pm0.06$& $72.9\pm25.9$& 2.68& B$_2$\\
90105-01-03-01 & 15/05/04& 7152&3023&Ch4E5& $4.944\pm0.009$&  $7.1\pm0.2$& 1.82& &$-0.77\pm0.09$& $34.7\pm10.9$& 2.42& B$_1$\\
90105-07-01-00 & 12/04/05& 6464&2098&Ch4E5& $4.018\pm0.005$& $10.4\pm0.2$& 1.62& &$-0.82\pm0.05$& $21.4\pm 2.3$& 0.95& B$_1$\\
90105-07-02-00 & 13/04/05& 6368&2123&Ch4E5& $3.890\pm0.006$& $11.1\pm0.2$& 1.99& &$-0.87\pm0.07$& $18.8\pm 2.6$& 1.88& B$_1$\\
90701-01-19-00 & 28/07/04& 6416&1200&Ch6E5& $2.116\pm0.002$& $14.5\pm0.4$& 1.89& &$-0.66\pm0.07$& $41.5\pm 9.2$& 3.05& B$_2$\\
91701-01-55-00 & 02/05/07& 9584&1091&Ch6E5& $1.986\pm0.003$& $15.1\pm0.6$& 2.69& &$-0.47\pm0.07$& $74.9\pm25.5$& 0.94& B$_2$\\
92702-01-09-00 & 04/05/06& 5136&1071&Ch6E5& $3.817\pm0.005$& $11.5\pm0.2$& 1.31& &$-0.99\pm0.11$& $14.2\pm 2.4$& 1.62& B$_2$\\
\hline
\end{longtable}

$^a$: The lengths of the good time intervals; \\
$^b$: ChID represent the definition of PCA energy bands for light curve extraction listed on Table S1; \\
$^c$: The LFQPO amplitude spectrum is fitted by a power-law with an exponential cutoff; \\
$^d$: No cutoff is detected in some observations at least up to $\sim100$ keV; \\
$^e$: In the LFQPO frequency-hardness diagram, the points follow two obviously separated branches which are labeled as "B$_1$" and "B$_2$", respectively.\\

\label{lastpage}

\end{document}